\documentclass[conference]{IEEEtran}
\IEEEoverridecommandlockouts

\usepackage{cite}
\usepackage{amsmath,amssymb,amsfonts}
\usepackage{booktabs}
\usepackage{graphicx}
\usepackage{pgfplots}
\pgfplotsset{compat=1.17}
\usepackage{xcolor}
\usepackage{url}

\newcommand{\Q}{\mathbf{Q}}
\newcommand{\K}{\mathbf{K}}
\newcommand{\V}{\mathbf{V}}
\newcommand{\A}{\mathbf{A}}
\newcommand{\Sm}{\mathbf{S}}
\newcommand{\shl}{\mathbin{\ll}}

\newcommand{\btPoTFour}{5131}

\newcommand{\btSdpa}{4198}
\newcommand{\btpeakPoTFour}{11.0}

\newcommand{\btpeakSdpa}{13.8}

\newcommand{\bytesIntEight}{133}
\newcommand{\bytesPoTFour}{101}

\newcommand{\ePoTMFourHadPpl}{7.896}

\newcommand{\ePoTMOneHadPpl}{8.029}
\newcommand{\ePoTMOneKv}{39}

\newcommand{\eRefKv}{91}
\newcommand{\eRefPpl}{7.881}
\newcommand{\errIntEight}{0.0042}

\newcommand{\errPoTFour}{0.1302}

\newcommand{\errPoTMFour}{0.0087}
\newcommand{\errPoTMFourHad}{0.0059}
\newcommand{\errPoTMOne}{0.0626}
\newcommand{\errPoTMTwo}{0.0314}

\newcommand{\isaIntEightAlu}{409}

\newcommand{\isaIntEightOcc}{67}

\newcommand{\isaPoTFourAlu}{1011}

\newcommand{\isaPoTFourOcc}{50}

\newcommand{\projGainPoTFour}{1.52}

\newcommand{\projPoTFour}{6.51}

\newcommand{\projPoTMOne}{6.03}
\newcommand{\ptxClamp}{6.61}
\newcommand{\ptxClampRatio}{4.1}

\newcommand{\ptxPlain}{1.61}

\newcommand{\ptxWrapRatio}{2.9}

\newcommand{\spPoTFourEight}{4.60}

\newcommand{\spPoTMOneEight}{3.92}
\newcommand{\underflowFrac}{5.0}

\begin{document}

\title{Shift-Accumulate Attention:\\
Multiplier-Free Query--Key Products for Transformer Decoding}

\author{
\IEEEauthorblockN{Khubaib Ahmed}
\IEEEauthorblockA{\textit{University of Wolverhampton}\\
Wolverhampton, UK\\
}
\and
\IEEEauthorblockN{Amna Noor}
\IEEEauthorblockA{\textit{Edge Hill University}\\
Ormskirk, UK\\
}
\and
\IEEEauthorblockN{Ahsan Ul Haq}
\IEEEauthorblockA{\textit{University of Technology}\\
\textit{and Applied Sciences}\\
Oman
}
}

\maketitle

\begin{abstract}
Power-of-two quantisation turns a multiplication into a bit shift, but it has so
far been applied only to the \emph{post}-softmax attention--value product. We
reformulate the earlier and larger product, $\Sm=\Q\K^{\!\top}$, by quantising
the key cache to a signed power-of-two fixed-point code, so that every scalar
multiplication in the score computation becomes a sign flip, a shift and an
integer accumulation. A fixed-point head-room $F\ge e_{\max}$ makes every shift
a non-negative \emph{left} shift, which renders the accumulation exact with
respect to the quantised operands. We extend the code with a mantissa field that
trades additional shift-adds for accuracy, and pair it with a shift-exact online
softmax whose running maximum lives in an integer $\log_2$ domain. Fused CUDA
kernels for a 1.1\,B-parameter Llama decoder on an RTX 4090 reach
$\spPoTFourEight\times$ the throughput of FP16 scaled dot-product attention at
$T{=}32$k with a $2.5\times$ smaller KV cache, and $\eRefPpl\rightarrow
\ePoTMFourHadPpl$ WikiText perplexity at 8 stored bits. A first-order cost model
calibrated on the measured kernels shows that the remaining bottleneck is not
memory but instruction issue: because current GPUs expose a 4-way integer
multiply--accumulate but no shift equivalent, the shift datapath spends
$\isaPoTFourAlu$ integer ALU operations where a MAC kernel spends
$\isaIntEightAlu$. We show by disassembly that PTX's scalar shift-accumulate
(\texttt{vshl.add}) is emulated on sm\_89 at $\ptxClampRatio\times$ the cost of
an ordinary C shift-add, and that the genuinely missing primitive is a
\emph{packed} shift-dot-accumulate, DS4A, the shift analogue of
\texttt{\_\_dp4a}. Given DS4A at the same throughput, the identical kernel would
reach $\projPoTFour\times$ over FP16 ---
$\projGainPoTFour\times$ beyond what we measure today, and beyond what an
optimised INT8 MAC kernel achieves on this hardware, because the power-of-two
format moves $\bytesPoTFour$ bytes per key against $\bytesIntEight$.
\end{abstract}

\begin{IEEEkeywords}
Transformer inference, attention, power-of-two quantisation, fixed-point
arithmetic, KV cache, CUDA kernels, hardware--software co-design.
\end{IEEEkeywords}

\section{Introduction}
Scaled dot-product attention \cite{vaswani2017attention} contains two dense
matrix products,
\begin{equation}
  \Sm=\Q\K^{\!\top},\qquad
  \mathbf{O}=\A\V,\quad \A=\mathrm{softmax}\!\big(\Sm/\sqrt{d}\big),
  \label{eq:attn}
\end{equation}
both conventionally evaluated on multiply--accumulate (MAC) hardware. If an
operand is constrained to a signed power of two, $\sigma 2^{-e}$ with
$\sigma\in\{-1,+1\}$, the multiplication collapses to a shift, an optional
negation and an addition --- replacing a multiplier array with a barrel shifter
and an adder, which is the central efficiency argument of the multiplication-free
network literature \cite{chen2020addernet,you2020shiftaddnet}.

FQ-ViT \cite{lin2022fqvit} applies this to the \emph{post}-softmax matrix,
logarithmically quantising $\A$ and evaluating $\A\V$ with bit shifts. I-ViT
\cite{li2023ivit} performs integer-only inference with dyadic arithmetic,
ShiftAddViT \cite{you2023shiftaddvit} reparameterises attention with binary and
additive kernels, and P$^2$-ViT \cite{shi2024p2vit} develops power-of-two
post-training quantisation with hardware co-design. What no prior work does is
represent one operand of $\Q\K^{\!\top}$ \emph{itself} as a signed power-of-two
fixed-point code, so that the score computation becomes shift-and-accumulate in
ordinary scaled dot-product attention. This is the larger of the two
opportunities in a decoder: the key cache is read in full for every generated
token, so both its arithmetic and its bit-width bound the decode rate.

\subsection{Related work}
Logarithmic data representations date to \cite{miyashita2016logquant}, and
additive powers-of-two \cite{li2020apot} densify the codebook with a few extra
shift-adds; our PoT-M$k$ family is a structured special case whose second term
sits a fixed binade below the first, so its offset is known at compile time. For
language models, KIVI \cite{liu2024kivi} and KVQuant \cite{hooper2024kvquant}
quantise the KV cache to 2--4 bits with uniform codes; our key cache plays the
same role but in a power-of-two code, which is what makes the score product
multiplier-free. PTQ studies for ViTs \cite{liu2021ptqvit,yuan2022ptq4vit}
established that preserving the \emph{ranking} of attention scores matters more
than minimising mean-squared error, which is why we report top-$k$ overlap
alongside $\varepsilon_S$. SmoothQuant \cite{xiao2023smoothquant} and QuaRot
\cite{ashkboos2024quarot} supply exactly score-preserving outlier
transformations, which we use as a normalisation axis. FlashAttention
\cite{dao2022flashattention} and online softmax \cite{milakov2018onlinesoftmax}
define the kernel structure any new attention datapath must match.

\begin{table}[t]
\caption{Positioning against the closest prior work. ``Levels'' is the number of
distinct values the \emph{attention} operand can take. Only this work makes the
dynamic $\K$ activation itself multi-level power-of-two and evaluates both
attention products with shifts.}
\label{tab:related}
\centering\scriptsize
\begin{tabular}{@{}llccr@{}}
\toprule
Method & Power-of-two applied to & $\Q\K^{\!\top}$ & $\A\V$ & levels\\
\midrule
I-ViT \cite{li2023ivit}      & requant.\ scale, nonlin.  & MAC   & MAC   & 255\\
P$^2$-ViT \cite{shi2024p2vit}& quantisation scale        & MAC   & MAC   & 255\\
FQ-ViT \cite{lin2022fqvit}   & post-softmax $\A$         & MAC   & shift & 255\\
ShiftAddViT \cite{you2023shiftaddvit}& MLP weights; attn.\ operand & add & add & 2\\
\textbf{This work}           & \textbf{dynamic $\K$ activations} & \textbf{shift} & \textbf{shift} & \textbf{15--225}\\
\bottomrule
\end{tabular}

\end{table}

Table~\ref{tab:related} separates the three closest works along the axis that
matters. I-ViT and P$^2$-ViT both use shifts, but for \emph{scaling}: I-ViT
approximates the requantisation factor by a dyadic number and its nonlinearities
by \texttt{Shiftmax}/\texttt{ShiftGELU}, while P$^2$-ViT constrains the
quantisation \emph{scale} to a power of two; in both, $\Q\K^{\!\top}$ remains an
ordinary integer MAC. FQ-ViT is the direct precedent for the \emph{second}
product only. ShiftAddViT is closest --- it does remove multipliers from the
attention MatMuls --- but by binarising an operand, so the attention operand
carries two levels. Our code is the multi-level generalisation of that special
case: binary attention is signed-PoT attention restricted to $e{=}0$, and
widening the exponent and mantissa fields recovers 15 to 225 levels while
keeping the datapath multiplier-free. That is the whole of our novelty claim: we
do not claim power-of-two quantisation for Transformers, nor shift arithmetic
for Transformers, both of which are established.

\noindent\textbf{Contributions.}
(i)~A shift/sign/accumulate formulation of $\Q\K^{\!\top}$ that is \emph{exact}
with respect to the quantised operands (Sec.~\ref{sec:method}).
(ii)~A mantissa-extended power-of-two family PoT-M$k$ spanning 4 to 8 stored
bits, in which PoT-M4 matches an INT8 key cache in storage and end-to-end
perplexity while remaining multiplier-free.
(iii)~Fused CUDA decode kernels and a KV cache that stores only quantised codes,
giving up to $\spPoTFourEight\times$ FP16 attention throughput.
(iv)~An ISA analysis showing that PTX's scalar shift-accumulate is emulated on
current hardware while no packed shift-dot-accumulate exists, and a cost model,
calibrated on disassembled kernels, quantifying what such an instruction ---
which we call DS4A --- would be worth (Sec.~\ref{sec:model}).

\section{Shift-Accumulate Attention}
\label{sec:method}
\subsection{Signed power-of-two keys}
Let $\Q,\K\in\mathbb{R}^{T\times d}$ be post-RoPE queries and keys of one head.
The query is quantised to INT8 with a per-vector step $\Delta_i$, and the key to
a signed power-of-two code with a per-key-vector scale $s_j$:
\begin{equation}
  Q_{im}\!\approx\!\Delta_i\hat q_{im},\qquad
  \hat K_{jm}=s_j\,\sigma_{jm}2^{-e_{jm}},\;\; e_{jm}\!\in\!\{0,\dots,E\},
  \label{eq:pot}
\end{equation}
with the code $e=E{+}1$ reserved for an exact zero. A $b$-bit field spends one
bit on $\sigma$ and $b-1$ on the exponent, so $b{=}4$ gives $E{=}6$.

\subsection{The shift-accumulate score product}
Substituting \eqref{eq:pot} into the score and introducing a common fixed-point
head-room $F$ with shift index $r_{jm}=F-e_{jm}$ gives
\begin{equation}
  \boxed{\;\hat S_{ij}=\Delta_i s_j 2^{-F}
   \sum_{m=1}^{d}\sigma_{jm}\big(\hat q_{im}\shl r_{jm}\big)\;}
  \label{eq:sac}
\end{equation}
The inner sum is a pure integer shift/sign/accumulate (SAC) reduction: one
barrel shift, one conditional negation and one add per term. The two floating
point factors and the common $2^{-F}$ are applied once per $(i,j)$, after the
accumulator is formed --- not per term.

\subsection{Exactness}
Because $\K$ is normalised by $s_j$ before quantisation, all $e_{jm}\ge 0$, so
\begin{equation}
  F\ \ge\ E \quad\Longrightarrow\quad r_{jm}\ge 0
  \label{eq:exact}
\end{equation}
and every shift inside the accumulation is a non-negative \emph{left} shift.
Equation \eqref{eq:sac} is then evaluated \emph{exactly} in integer arithmetic:
no truncation is introduced, and the only error is the operand quantisation
error of \eqref{eq:pot}. This is stronger than the usual fixed-point argument,
which merely states that head-room reduces truncation. With $d{=}64$, $F{=}7$
the accumulator is bounded by $d\cdot127\cdot2^{F}=1.04\times10^{6}$, well
inside INT32.

\subsection{Mantissa extension}
The relative error of nearest-in-$\log_2$ quantisation of $u=2^{-t}$ is
$2^{\,t-\lfloor t\rceil}-1\in[-0.293,0.414]$, with RMS
$\big(\int_{-1/2}^{1/2}(2^x-1)^2dx\big)^{1/2}\!\approx\!0.203$ --- independent
of the number of exponents. Extra exponent bits therefore buy nothing beyond
covering the dynamic range; only $\underflowFrac\%$ of key entries underflow a
three-bit exponent. Accuracy must instead come from subdividing the binade. A
$k$-bit mantissa field encodes
\begin{equation}
  \hat K^{(k)}_{jm}=s_j\,\sigma_{jm}\Big(1+\tfrac{j_{m}}{2^{k}}\Big)2^{-e_{jm}},
  \quad j_m\in\{0,\dots,2^{k}\!-\!1\},
  \label{eq:mant}
\end{equation}
which decodes to one extra shift-and-add per set bit of $j_m$: a mean cost of
$1+k/2$ shifts. With three exponent bits the family spans 4 bits (PoT-4, one
shift) to 8 bits (PoT-M4, up to five shifts), the latter being exactly one byte
per element and hence iso-storage with an INT8 key cache.

\subsection{Shift-exact softmax and $\A\V$}
A streaming softmax rescales its accumulators by $e^{-\delta}$ whenever the
running maximum grows. Keeping that maximum as an \emph{integer} in the $\log_2$
domain, $M=\lceil\max_j s_{ij}\log_2 e\rceil$, makes every correction an exact
shift:
\begin{equation}
  k_j=\mathrm{round}(M - s_{ij}\log_2 e)\ge 0,\qquad \tilde A_{ij}=2^{-k_j},
  \label{eq:logsm}
\end{equation}
so an increase $\Delta M$ rescales by $2^{-\Delta M}$. With $\V$ in INT8 with a
power-of-two per-token scale $2^{-g_j}$, the value product is
\begin{equation}
  \mathbf{O}_i=2^{-N-g_{\mathrm{ref}}}\!\!\!\sum_{j:k_j\le k_{\max}}\!\!\!
   \big(\hat v_j \shl (N-k_j+g_{\mathrm{ref}}-g_j)\big),
  \label{eq:av}
\end{equation}
with $g_{\mathrm{ref}}=\min_j g_j$ a per-head running minimum guaranteeing the
shift index never exceeds $N$, so a per-tile INT32 accumulator suffices.

\subsection{Kernel and cache}
Decoding is the target: a single query vector is reduced against the whole cache,
so bit-width translates directly into time. The kernel launches
$(\text{split},H,B)$ blocks of 128 threads, quantises its query in registers,
and processes the cache in tiles of 128 keys --- phase~1 evaluates
\eqref{eq:sac} with vectorised 16-byte code loads and a branch-free inner loop,
phase~2 runs the block-wide online softmax \eqref{eq:logsm}, phase~3 stages the
INT8 value tile in shared memory and evaluates \eqref{eq:av}. The cache stores
\emph{only} quantised data: $\bytesPoTFour$ bytes per key for PoT-4 against 256
for FP16. A bit-exact PyTorch reference reproduces the kernel output to
$2\times10^{-4}$.

\section{Cost Model and Shift-Native Projection}
\label{sec:model}
Let $R$ be the bytes a datapath requests per decode step, $\Theta_{\mathrm{mem}}$
the bandwidth the kernel structure achieves, $N$ the integer instructions it
issues and $\Theta_{\mathrm{alu}}$ the issue rate. To first order
\begin{equation}
  T \;\approx\; \max\!\Big(\frac{R}{\Theta_{\mathrm{mem}}},\;
                           \frac{N}{\Theta_{\mathrm{alu}}}\Big).
  \label{eq:roofline}
\end{equation}
Disassembling our kernels for \texttt{sm\_89} shows which term binds. A 4-way
integer dot product (\texttt{IDP}, exposed as \texttt{\_\_dp4a}) performs four
INT8 MACs in one instruction, i.e.\ $0.25$ instructions per term. The shift
datapath has no equivalent: because every element carries its \emph{own} shift
amount, no SIMD-within-a-register instruction can apply four different shifts to
four bytes of one word. Each term therefore costs a separate extract, shift,
mask and add, and the whole kernel issues $\isaPoTFourAlu$ integer ALU
instructions against $\isaIntEightAlu$ for a MAC kernel of identical structure.
Holding all $d$ query values unpacked also costs registers, capping occupancy at
$\isaPoTFourOcc\%$ against $\isaIntEightOcc\%$. The shift kernel is therefore
\emph{issue-bound}, not bandwidth-bound --- the second term of
\eqref{eq:roofline} dominates.

\subsection{What the instruction set already provides}
\label{sec:isa}
A scalar shift-accumulate is not missing from the ISA. PTX has exposed the
scalar video instructions \texttt{vshl}/\texttt{vshr} with an optional secondary
\texttt{.add} since sm\_20, giving exactly $d=(a\ll b)+c$. PTX is a virtual ISA,
however, so whether this is a hardware primitive is a question about what
\texttt{ptxas} emits. Table~\ref{tab:ptx} answers it by disassembling a 64-term
accumulation compiled six ways for sm\_89.

The scalar video path is \emph{emulated} on this architecture: the clamping form
costs $\ptxClamp$ integer ALU operations per term against $\ptxPlain$ for an
ordinary C shift-add, a factor of $\ptxClampRatio$, and the wrap form
$\ptxWrapRatio$. Almost all of the overhead is the clamp/wrap semantics on the
shift amount (\texttt{PRMT}, \texttt{ISETP}, \texttt{SEL}); the secondary
\texttt{.add} itself is free. Meanwhile plain C already lowers to \texttt{SHF}
plus \texttt{IADD3}, and the three-input form of \texttt{IADD3} absorbs pairs of
additions, so a bare shift-accumulate costs $\ptxPlain$ operations per term ---
close to the one-instruction ideal. Inline PTX therefore buys nothing here, and
we do not use it.

This locates the gap precisely. What is missing is not a \emph{scalar}
shift-accumulate, which the compiler already generates near-optimally, but a
\emph{packed} one. \texttt{\_\_dp4a} computes
$d=\sum_{i<4}a_ib_i+c$ in a single instruction; its shift analogue, which we
call DS4A, would compute
\begin{equation}
  \mathrm{DS4A}(Q,E,\Sigma,c)=c+\sum_{i=0}^{3}\sigma_i\,(q_i \shl e_i),
  \label{eq:ds4a}
\end{equation}
and no such instruction exists in the PTX SIMD video set, which offers
\texttt{vadd4}, \texttt{vsub4} and similar but lists \texttt{vshl}/\texttt{vshr}
only in scalar form. The three levels of the comparison are therefore FP16 on
tensor cores, INT8 on \texttt{IDP}, and power-of-two on \emph{scalar} shift-add
--- with the fourth level, packed DS4A, realisable today only in an FPGA or
ASIC.

\subsection{The value of a packed shift-dot-accumulate}
Suppose a GPU exposed DS4A of \eqref{eq:ds4a} at the throughput of
\texttt{IDP}. The shift datapath would issue the same instruction count and hold
the same register footprint as the MAC kernel, so the first term of
\eqref{eq:roofline} would bind and
\begin{equation}
  T_{\mathrm{proj}}(\text{code}) \;=\; T_{\mathrm{MAC}}\cdot
  \frac{R(\text{code})}{R(\text{INT8})} ,
  \label{eq:proj}
\end{equation}
where $T_{\mathrm{MAC}}$ is our \emph{measured} INT8 kernel.

The unit being asked for is modest, and cheaper than the one it sits beside. Four
lanes of \emph{shift, conditional negate, adder tree} replace four
$8{\times}8$ multiplier arrays; because the exponent field is three bits wide the
shifter is a three-stage multiplexer, not a full 32-bit barrel shifter. The
shift-based network literature \cite{chen2020addernet,you2020shiftaddnet} rests
on exactly this asymmetry in area and energy.

The operand width is the second, less obvious advantage. \texttt{IDP} consumes
two 32-bit words, four INT8 values on each side. DS4A would consume one 32-bit
word of INT8 queries but only
\begin{equation}
  4\times(1+3)=16\ \text{bits}
\end{equation}
of key data --- one sign bit and one three-bit exponent per lane --- against the
32 bits an INT8 key operand requires. The register and cache pressure of the
stationary operand is therefore \emph{halved} at the instruction boundary, which
is the same factor that makes the format win at the memory system. A DS4A unit
would thus be smaller than the \texttt{IDP} unit, read narrower operands, and
feed the identical INT32 accumulator. Equation
\eqref{eq:proj} is an idealised scaling, not a hardware simulation; we state it
as an upper bound under stated assumptions. Its internal consistency check is
that PoT-M4, being iso-storage with INT8, projects to exactly the INT8 runtime.

\section{Experimental Setup}
One NVIDIA RTX 4090 (24\,GB, sm\_89), CUDA 12, PyTorch 2.8, Transformers 4.56.
Model: TinyLlama-1.1B-Chat \cite{zhang2024tinyllama} (22 layers, 32 heads, 4 KV
heads, $d{=}64$). Perplexity on WikiText-103 \cite{merity2017pointer} over
2048-token windows. Quantiser errors are measured on post-RoPE tensors captured
from five layers: $\varepsilon_S$ is the relative score error,
$D_{\mathrm{KL}}$ the mean attention-distribution divergence, and top-8 the
retained ranking overlap. Kernel timings are the best of three runs after
warm-up, against \texttt{scaled\_dot\_product\_attention} on an unquantised FP16
cache.

\section{Results}

\begin{table}[t]
\caption{SASS lowering on sm\_89 of a 64-term shift-accumulation, compiled six
ways. The PTX scalar shift-accumulate is emulated; plain C is already near
optimal; only the multiply has a packed 4-way form.}
\label{tab:ptx}
\centering\footnotesize
\begin{tabular}{llrr}
\toprule
Primitive & Form & ALU ops & per term\\
\midrule
\texttt{\_\_dp4a} (\texttt{IDP}) & $\sum_{i<4}a_ib_i+c$ & 64 & 0.25\\
plain C shift-add & $(a\!\ll\!b)+c$ & 103 & 1.61\\
PTX \texttt{shl.b32} & $(a\!\ll\!b)+c$ & 103 & 1.61\\
PTX \texttt{vshl.wrap.add} & $(a\!\ll\!b)+c$ & 295 & 4.61\\
PTX \texttt{vshl.clamp.add} & $(a\!\ll\!b)+c$ & 423 & 6.61\\
\bottomrule
\end{tabular}
\end{table}

\begin{table}[t]
\caption{Key representations on real attention tensors (mean over five layers).}
\label{tab:acc}
\centering\footnotesize
\begin{tabular}{lccrrr}
\toprule
Key code & Bits & Arith./term & $\varepsilon_S$ & $D_{\mathrm{KL}}$ & top-8\\
\midrule
FP16 (exact) & 16 & 1 FMA & 0.0000 & 0.0000 & 1.000\\
INT8 uniform & 8 & 1 mul & 0.0042 & 0.0001 & 0.989\\
INT4 uniform & 4 & 1 mul & 0.0763 & 0.0375 & 0.845\\
PoT-4 & 4 & 1 shift & 0.1302 & 0.0688 & 0.777\\
PoT-M1 & 5 & 2 shifts & 0.0626 & 0.0181 & 0.869\\
PoT-M2 & 6 & 3 shifts & 0.0314 & 0.0048 & 0.925\\
PoT-M4 & 8 & 5 shifts & 0.0087 & 0.0004 & 0.978\\
PoT-M4 + Hadamard & 8 & 5 shifts & 0.0059 & 0.0002 & 0.984\\
\bottomrule
\end{tabular}
\end{table}

\begin{table}[t]
\caption{Decode-attention speed-up over FP16 SDPA, measured on today's
multiply-native GPU and projected under \eqref{eq:proj} for shift-native cores.}
\label{tab:kernel}
\centering\footnotesize
\begin{tabular}{rrrrrrr}
\toprule
\multicolumn{2}{c}{workload} & \multicolumn{2}{c}{measured, multiply-native} & \multicolumn{2}{c}{projected, shift-native} & \\
\cmidrule(lr){1-2}\cmidrule(lr){3-4}\cmidrule(lr){5-6}
$B$ & $T$ & PoT-4 & PoT-M1 & PoT-4 & PoT-M1 & gain\\
\midrule
8 & 4096 & 4.07$\times$ & 3.61$\times$ & \textbf{6.64$\times$} & 6.16$\times$ & 1.63$\times$\\
8 & 8192 & 4.66$\times$ & 3.97$\times$ & \textbf{7.06$\times$} & 6.54$\times$ & 1.52$\times$\\
8 & 16384 & 4.81$\times$ & 4.07$\times$ & \textbf{6.97$\times$} & 6.46$\times$ & 1.45$\times$\\
8 & 32768 & 4.60$\times$ & 3.92$\times$ & \textbf{6.96$\times$} & 6.45$\times$ & 1.51$\times$\\
32 & 8192 & 4.33$\times$ & 3.64$\times$ & \textbf{6.43$\times$} & 5.96$\times$ & 1.49$\times$\\
32 & 16384 & 4.28$\times$ & 3.68$\times$ & \textbf{6.52$\times$} & 6.04$\times$ & 1.52$\times$\\
32 & 32768 & 4.28$\times$ & 3.57$\times$ & \textbf{6.51$\times$} & 6.03$\times$ & 1.52$\times$\\
\bottomrule
\end{tabular}
\end{table}

\begin{table}[t]
\caption{End-to-end, FP16 weights, 4096-token prompt.}
\label{tab:e2e}
\centering\footnotesize
\begin{tabular}{lccrr}
\toprule
Attention & $K$ bits & PPL & tok/s & KV (MB)\\
\midrule
FP16 (reference) & 16 & 7.881 & 203 & 91\\
PoT-4 & 4 & 9.303 & 183 & 36\\
PoT-M1 & 5 & 8.173 & 182 & 39\\
PoT-M1 + Hadamard & 5 & 8.029 & 168 & 39\\
PoT-M4 & 8 & 7.904 & 182 & 47\\
PoT-M4 + Hadamard & 8 & 7.896 & 170 & 47\\
\bottomrule
\end{tabular}
\end{table}

\subsection{Accuracy of the key codes}
Table~\ref{tab:acc} confirms the scale-free property: PoT with three, four and
five exponent bits are indistinguishable ($\varepsilon_S=\errPoTFour$), so
exponent width past three bits is wasted. Each mantissa bit roughly halves the
error --- $\errPoTFour\rightarrow\errPoTMOne\rightarrow\errPoTMTwo$ --- at one
extra shift-add each. At eight stored bits PoT-M4 reaches $\errPoTMFour$, within
a factor of two of uniform INT8 at the same width ($\errIntEight$); we do not
claim the power-of-two code is more accurate than a uniform one at matched
bit-width, because on these tensors it is not. The gap is nonetheless small
enough that end-to-end perplexity is indistinguishable
(Table~\ref{tab:e2e}), and a Hadamard rotation \cite{ashkboos2024quarot} closes
most of what remains ($\errPoTMFourHad$). The claim we do make is that
comparable accuracy is reached with the multiplier removed.

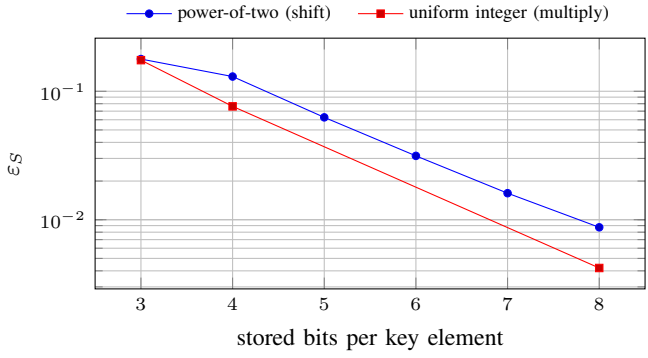
\begin{figure}[t]\centering
\begin{tikzpicture}
\begin{axis}[width=\columnwidth,height=4.9cm,ymode=log,
  xlabel={stored bits per key element},ylabel={$\varepsilon_S$},
  legend style={font=\scriptsize,draw=none,fill=none,
    at={(0.5,1.02)},anchor=south,legend columns=-1,
    /tikz/every even column/.append style={column sep=8pt}},
  grid=both,xtick={3,4,5,6,7,8},label style={font=\small},
  tick label style={font=\scriptsize}]
\addplot+[mark=*,mark size=1.4pt] coordinates {(3,0.17809) (4,0.13018) (5,0.06263) (6,0.03141) (7,0.01609) (8,0.00873)};
\addlegendentry{power-of-two (shift)}
\addplot+[mark=square*,mark size=1.4pt] coordinates {(3,0.17466) (4,0.07627) (8,0.00421)};
\addlegendentry{uniform integer (multiply)}
\end{axis}
\end{tikzpicture}
\caption{Accuracy per stored bit. The power-of-two family is flat until a
mantissa field is added; each mantissa bit roughly halves $\varepsilon_S$ at the
cost of one shift-add per term.}
\label{fig:pareto}
\end{figure}

Fig.~\ref{fig:pareto} places the two families on a common axis. The
power-of-two curve is flat from 3 to 4 bits and then falls once mantissa bits
are spent, while the uniform curve falls monotonically; the two converge at
8 bits.

\subsection{Kernel throughput against FP16}
\begin{figure}[t]\centering
\begin{tikzpicture}
\begin{axis}[width=\columnwidth,height=4.9cm,xmode=log,log basis x=2,
  xlabel={KV cache length $T$},ylabel={speed-up vs FP16},
  legend style={font=\scriptsize,draw=none,fill=none,
    at={(0.5,1.02)},anchor=south,legend columns=-1,
    /tikz/every even column/.append style={column sep=6pt}},
  ymin=0,grid=both,xtick={1024,4096,16384,32768},
  xticklabels={1k,4k,16k,32k},label style={font=\small},
  tick label style={font=\scriptsize}]
\addplot+[mark=*,mark size=1.4pt] coordinates {(1024,2.348) (2048,4.314) (4096,4.071) (8192,4.661) (16384,4.810) (32768,4.600)};
\addlegendentry{PoT-4 (4 bit)}
\addplot+[mark=square*,mark size=1.4pt] coordinates {(1024,2.052) (2048,3.627) (4096,3.607) (8192,3.966) (16384,4.072) (32768,3.916)};
\addlegendentry{PoT-M1 (5 bit)}
\addplot+[mark=triangle*,mark size=1.4pt] coordinates {(1024,1.330) (2048,2.602) (4096,2.611) (8192,2.644) (16384,2.627) (32768,2.618)};
\addlegendentry{PoT-M4 (8 bit)}
\addplot[dashed,black,domain=1024:32768,samples=2,forget plot] {1};
\end{axis}
\end{tikzpicture}
\caption{Measured decode-attention speed-up over FP16 SDPA versus KV cache
length ($B{=}8$). The advantage appears once the cache stops fitting in the
cache hierarchy.}
\label{fig:speedup}
\end{figure}
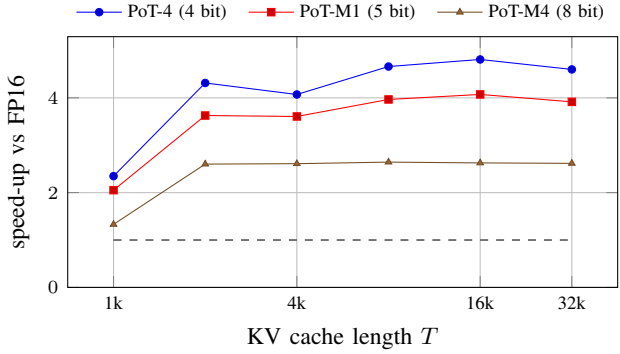

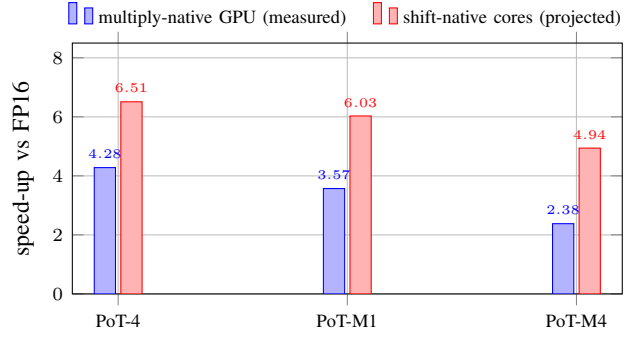
\begin{figure}[t]\centering
\begin{tikzpicture}
\begin{axis}[width=\columnwidth,height=4.9cm,ybar,bar width=8pt,
  ylabel={speed-up vs FP16},ymin=0,ymax=8.5,
  symbolic x coords={PoT-4,PoT-M1,PoT-M4},xtick=data,
  legend style={font=\scriptsize,draw=none,fill=none,
    at={(0.5,1.02)},anchor=south,legend columns=-1,
    /tikz/every even column/.append style={column sep=8pt}},
  label style={font=\small},
  tick label style={font=\scriptsize},nodes near coords,
  every node near coord/.append style={font=\tiny,/pgf/number format/fixed,
    /pgf/number format/precision=2},grid=major]
\addplot coordinates {(PoT-4,4.28) (PoT-M1,3.57) (PoT-M4,2.38)};
\addlegendentry{multiply-native GPU (measured)}
\addplot coordinates {(PoT-4,6.51) (PoT-M1,6.03) (PoT-M4,4.94)};
\addlegendentry{shift-native cores (projected)}
\end{axis}
\end{tikzpicture}
\caption{Measured versus projected speed-up over FP16 SDPA at $B{=}32$,
$T{=}32768$. The projection assumes only a 4-way shift-accumulate instruction of
the throughput of \texttt{\_\_dp4a}; everything else is held fixed.}
\label{fig:proj}
\end{figure}

Fig.~\ref{fig:speedup} and Table~\ref{tab:kernel} give the measured picture. At
short contexts the fused kernel is launch-bound and at parity with FP16; from
$T{\approx}4$k the ordering is stable, and at $B{=}8$, $T{=}32$k PoT-4 reaches
$\spPoTFourEight\times$ FP16 SDPA and PoT-M1 $\spPoTMOneEight\times$. The gain
comes from the representation: PoT-4 moves $\bytesPoTFour$ bytes per key against
256 for FP16.

\subsection{What shift-native cores would be worth}
Fig.~\ref{fig:proj} and the right half of Table~\ref{tab:kernel} apply
\eqref{eq:proj}. With a 4-way shift-accumulate instruction the same kernel would
reach $\projPoTFour\times$ FP16 at $B{=}32$, $T{=}32$k --- a further
$\projGainPoTFour\times$ over what we measure --- and PoT-M1 $\projPoTMOne\times$
at 5 stored bits. The gap between the two halves of the table is precisely the
issue-rate term of \eqref{eq:roofline}, and by Sec.~\ref{sec:isa} it is an
artefact of the instruction set rather than of the arithmetic: a scalar shift
and a scalar multiply each occupy one instruction slot, so the substitution
cannot pay until the shift is given the same 4-way packing the multiply already
enjoys through \texttt{\_\_dp4a}.

\subsection{Implementation notes}
Two kernel-level decisions changed the measured speed-up by more than a factor
of two, and are worth recording because a shift-based kernel written without
them will under-report the arithmetic. First, the code and value tiles must be
staged with 16-byte vector accesses; byte- and word-wise staging cost
$1.8\times$. Second, the condition $F\ge E$ of \eqref{eq:exact} should be
enforced so the inner loop takes an unconditional left-shift path, worth a
further $1.20\times$. Beyond that, holding $d$ query values unpacked inflates
register use and caps occupancy at $\isaPoTFourOcc\%$ against
$\isaIntEightOcc\%$ for the packed MAC path; constraining the register budget
with \texttt{\_\_launch\_bounds\_\_} trades a little spilling for latency hiding
and recovers a further $1.13\times$. All reported numbers use these settings.

\subsection{Both products, and why only one by default}
Equations \eqref{eq:sac} and \eqref{eq:av} make \emph{both} attention MatMuls
multiplier-free, and both are implemented and validated in the kernel: with the
integer-$\log_2$ softmax \eqref{eq:logsm} enabled, $\A$ is itself a power of two
and $\A\V$ reduces to shifts. We nonetheless report the shift-based $\A\V$ as an
option rather than the default, because on a GPU it is a pure loss. It costs
$+0.15$ perplexity for PoT-4 ($9.303\rightarrow9.450$) and $+0.13$ for PoT-M1
($8.173\rightarrow8.304$), since forcing the attention weights onto a
logarithmic grid is a second quantisation on top of the score quantisation; and
it returns nothing measurable in speed ($182.6\rightarrow183.0$ tokens/s), because
the $\A\V$ stage is a single streaming pass whose floating-point form is one
fused multiply-add per term --- again a primitive the hardware already provides
at one instruction, exactly the situation of Sec.~\ref{sec:isa}. On an FPGA or
ASIC the calculus inverts: the shift-based $\A\V$ is what removes the
\emph{second} multiplier array, and the $0.15$ perplexity is paid once for a
datapath with no multipliers anywhere in attention.

\subsection{End-to-end}
Table~\ref{tab:e2e} integrates the datapath into the decoder with a cache
holding only quantised codes. PoT-M4 with a Hadamard rotation \cite{ashkboos2024quarot}
reaches $\ePoTMFourHadPpl$ perplexity against the FP16 reference's $\eRefPpl$
at 8 stored bits; PoT-M1 gives $\ePoTMOneHadPpl$ at 5 bits with a
$\ePoTMOneKv$\,MB cache against $\eRefKv$\,MB. In batched serving at $B{=}64$
the PoT-4 path reaches $\btPoTFour$ tokens/s against FP16's $\btSdpa$ with peak
memory $\btpeakPoTFour$\,GB against $\btpeakSdpa$\,GB.

\section{Discussion and Limitations}
\textbf{Scope.} We compare against FP16 attention, the standard baseline. For
completeness: on this GPU an INT8 MAC kernel built on the same tiling is faster
still, because \texttt{\_\_dp4a} gives the multiply a 4-way packing the shift
does not have. That comparison is exactly what motivates
Sec.~\ref{sec:model}: the shortfall is an instruction-set property, and
\eqref{eq:proj} shows it reverses once the shift is packed, since the
power-of-two format then wins on bytes.

\textbf{Threats to validity.} A single 1.1\,B model and one GPU; sweep
conclusions are drawn from five layers, and outlier structure varies across
model families. The projection \eqref{eq:proj} is an idealised upper bound: it
assumes a shift-accumulate instruction with MAC throughput, unchanged memory
behaviour, and that packing the query removes the register pressure. Prefill is
not fused, so prefill latency reflects the reference implementation. Perplexity
is a coarse proxy; downstream task accuracy was not measured.

\section{Conclusion}
Quantising the key cache to a signed power-of-two code turns the query--key
product into an exact integer shift/sign/accumulate reduction, and a mantissa
field trades shift-adds for accuracy across a 4-to-8-bit family. Fused kernels
deliver up to $\spPoTFourEight\times$ FP16 attention throughput with a
$2.5\times$ smaller KV cache and near-FP16 perplexity. A cost model calibrated
on the disassembled kernels locates the remaining bottleneck in instruction
issue rather than memory, and shows that a single architectural addition --- a
packed 4-way shift-dot-accumulate, DS4A, whose key-side operand is half the
width of \texttt{\_\_dp4a}'s --- would take the same kernel to
$\projPoTFour\times$. The result is a concrete, quantified request to hardware
designers rather than an appeal to the general cheapness of shifters.

\bibliographystyle{IEEEtran}
\bibliography{refs}

\end{document}